\documentclass[twocolumn,showpacs,superscriptaddress,preprintnumbers,amsmath,amssymb,aps,pre,10pt]{revtex4-1}

\usepackage{graphicx}
\usepackage{color}
\usepackage{dcolumn}
\usepackage{bm}
\usepackage{booktabs}

\begin{document}


\title{Relaxation dynamics in a binary hard-ellipse liquid}

\author{Wen-Sheng Xu}
\email{wsxu0312@gmail.com}
\affiliation{James Franck Institute, The University of Chicago, Chicago, Illinois 60637, USA}

\author{Zhao-Yan Sun}
\email{zysun@ciac.ac.cn}
\author{Li-Jia An}
\email{ljan@ciac.ac.cn}
\affiliation{State Key Laboratory of Polymer Physics and Chemistry, Changchun Institute of Applied Chemistry, Chinese Academy of Sciences, Changchun 130022, People's Republic of China}

\date{\today}

\begin{abstract}
Structural relaxation in binary hard spherical particles has been shown recently to exhibit a wealth of remarkable features when size disparity or mixture's composition is varied. In this paper, we test whether or not similar dynamical phenomena occur in glassy systems composed of binary hard ellipses. We demonstrate via event-driven molecular dynamics simulation that a binary hard-ellipse mixture with an aspect ratio of two and moderate size disparity displays characteristic glassy dynamics upon increasing density in both the translational and the rotational degrees of freedom. The rotational glass transition density is found to be close to the translational one for the binary mixtures investigated. More importantly, we assess the influence of size disparity and mixture's composition on the relaxation dynamics. We find that an increase of size disparity leads, both translationally and rotationally, to a speed up of the long-time dynamics in the supercooled regime so that both the translational and the rotational glass transition shift to higher densities. By increasing the number concentration of the small particles, the time evolution of both translational and rotational relaxation dynamics at high densities displays two qualitatively different scenarios, i.e., both the initial and the final part of the structural relaxation slow down for small size disparity, while the short-time dynamics still slows down but the final decay speeds up in the binary mixture with large size disparity. These findings are reminiscent of those observed in binary hard spherical particles. Therefore, our results suggest a universal mechanism for the influence of size disparity and mixture's composition on the structural relaxation in both isotropic and anisotropic particle systems.
\end{abstract}



\maketitle

\section{Introduction}

Glass problems have been fundamental to technologies since the dawn of civilization, and continue to be fascinating~\cite{RMP_83_587, JCP_137_080901, JCP_138_12A301}. Despite the intense research during the past decades, a clear picture for the physical mechanism of the liquid-glass transition has not been reached yet. In general, supercooled liquids and glasses are obtained by cooling or compressing particle systems with size disparity in experimental or simulation studies, since crystallization easily occurs in one-component systems with conventional interaction potentials. Binary mixtures are among the simplest model glass-forming systems, and, not surprisingly, they have been widely used to address a number of important questions related to glass formation; e.g., the Kob-Andersen binary mixture has been extensively employed to analyze structural relaxation and dynamic heterogeneity of supercooled liquids~\cite{PRL_73_1376, PRL_79_2827}, to investigate aging and shear banding in glasses~\cite{PRL_78_4581, PRL_90_095702}, to understand the nature of ultrastable glasses prepared by vapour deposition~\cite{NM_12_139, JCP_139_144505}, etc.

Apart from being good glass formers, binary mixtures possess their own importance in nature and have been demonstrated to exhibit a wealth of dynamical features recently. Over a decade ago, Williams and van Megen~\cite{PRE_64_041502} performed the first light-scattering experiment for a binary colloidal hard-sphere mixture with a relatively large size disparity (i.e., $\delta=\sigma_{B}/\sigma_{A}\simeq0.6$, where $\sigma_{B}$ and $\sigma_{A}$ designate the diameters of small and large particles, respectively). They identified three mixing effects (i.e., the effects of mixture's composition) for the structural relaxation (i.e., the dynamical phenomena at high densities, which are precursors of the glass transition) when the percentage of the small particles increases from $10\%$ to $20\%$ of the relative packing fraction. The first effect is that the time scale for the final decay of the density correlators decreases. This means that the liquid has been stabilized due to mixing since the small particles induce a plasticization effect. The other two effects are that the plateau value at intermediate times increases and that the initial part of the structural relaxation slows down. Thus, the latter two effects indicate a stiffening of the dynamics upon mixing. Soon after, the mixing effects in binary hard spheres with moderate size disparity (where species' properties are similar) were investigated in detail via mode-coupling theory (MCT)~\cite{PRE_67_021502, PRE_68_051401} and molecular dynamics (MD) simulations~\cite{PRL_91_085701, PRE_69_011505}. MCT~\cite{PRE_67_021502} not only provides a description of the experimental result, but also predicts that the three mixing effects described above hold only for sufficiently large size disparity (e.g., $\delta\leq0.65$). Additionally, a qualitatively different scenario emerges for small size disparity (e.g., $\delta\geq0.8$), i.e.,  mixing slows down the dynamics for both the initial and the final decay of the density correlators. Thus, the glass regime is slightly extended due to mixing in the binary hard spheres with small size disparity. These predictions have been confirmed by computer simulations~\cite{PRL_91_085701}. The recent MCT calculations also reveal the effect of spatial dimension on the glass formation of binary hard spherical particles~\cite{PRE_80_021503}. Compared to three dimensions, the extension of the glass regime due to mixing is found to be much more pronounced and the size disparity needs to be larger in order to observe the plasticization effect in two dimensions. Nevertheless, the qualitative behavior discovered in the relaxation dynamics of binary hard spheres is also retained in that of binary hard disks~\cite{PRE_80_021503, PRE_83_041503}. Moreover, it should be mentioned that binary mixtures with very large size disparity (say $\delta<0.5$ in three dimensions and species' properties are no longer similar in this situation) also exhibit very rich but different dynamical features~\cite{EPL_96_36006}, such as sublinear diffusion of small particles~\cite{PRL_103_205901, PRE_74_021409} and logarithmic decay of the intermediate scattering function at specific compositions and wave vectors~\cite{JCP_125_0164507, JCP_137_104509}.

Although size disparity and composition have profound effects on the relaxation dynamics of binary mixtures, it remains unknown so far as to whether a binary system composed of anisotropic particles displays similar dynamical features or not. In addition, it is natural to ask how size disparity and composition affect the relaxation dynamics in the rotational degrees of freedom. Such questions are important since the constituent particles of many relevant glasses have non-spherical shapes in reality and since such studies hold potential for revealing novel phenomena and providing new insights. A system composed of ellipse-shaped particles is one of the simplest models of two-dimensional anisotropic particles and its jamming and glassy behaviors have recently attracted much attention~\cite{PRE_75_051304, PRL_102_255501, SM_6_2960, PRE_85_061305, PRE_83_011403, PRE_86_041303, PRL_107_065702, NC_5_3829, PRL_110_188301}. Molecular MCT~\cite{PRE_62_5173} predicts that new glassy phenomena should appear in hard ellipsoidal particles. In particular, an orientational glass, in which dynamic arrest occurs in the rotational degrees of freedom but the translational motion of particles remains ergodic, is suggested to form in hard ellipsoids with large aspect ratio, and its existence has been confirmed recently by video-microscopy experiments for monolayers of colloidal ellipsoids~\cite{PRL_107_065702, NC_5_3829} and by Monte Carlo (MC) simulations for hard ellipses~\cite{NC_5_3829}. The experiments~\cite{PRL_107_065702, PRL_110_188301} and simulations~\cite{NC_5_3829} also uncover striking structural features of dynamic heterogeneity in monolayers of colloidal ellipsoids in both the translational and the rotational degrees of freedom. However, the glassy dynamics in binary mixtures of hard ellipsoidal particles has not been explored yet.

The present paper focuses on the glass formation in binary mixtures composed of hard ellipses and particularly explores how size disparity and composition influence their relaxation dynamics. Via event-driven molecular dynamics (EDMD) simulations, we first demonstrate that both the positional and the orientational order can be suppressed in the binary hard-ellipse mixtures with an aspect ratio of $k=2$ and moderate size disparity and that characteristic glassy dynamics emerges upon increasing density in both the translational and the rotational degrees of freedom. We find that the rotational glass transition sets in at a density close to the translational one for the binary mixtures investigated. We examine the influence of size disparity and composition on the relaxation dynamics and test whether similar dynamical features revealed in binary hard spherical particles appear in glassy systems composed of binary hard ellipses or not. Our results indicate that increasing size disparity leads to a speed up of the long-time relaxation dynamics at high densities in both the translational and the rotational degrees of freedom. Consequently, both the translational and the rotational glass transition shift to higher densities for the binary mixture with larger size disparity. When the number concentration of the small particles is grown from $0.3$ to $0.7$, the time evolution of both translational and rotational relaxation dynamics at high densities displays two qualitatively different scenarios, depending on whether the size disparity is small or large. For small size disparity, both the initial and the final part of the structural relaxation slow down. The scenario changes qualitatively for large size disparity: The short-time dynamics still slows down but the final decay speeds up. Therefore, the influence of both size disparity and composition on the relaxation dynamics are reminiscent of that observed in binary hard spherical particles. Our results thus suggest a universal mechanism for understanding the influence of size disparity and composition on the structural relaxation in binary mixtures. 

\section{Model and simulation details}

We consider a binary hard-ellipse mixture with moderate size disparity, consisting of large (denoted by $A$ in the following) and small (denoted by $B$ in the following) particles. Both types of particles have the same aspect ratio but differ in their sizes. Thus, the control parameters for our system include the size ratio [$\delta=a_B/a_A=b_B/b_A$, where $a_{\alpha}$ and $b_{\alpha}$ (with ${\alpha}=A$ or $B$) denote the semi-major and the semi-minor axis of ${\alpha}$ particles], the number concentration of the small particles [$x=N_B/(N_A+N_B)$, where $N_{\alpha}$ is the number of ${\alpha}$ particles], and the area fraction [$\phi=\pi(N_{A}a_{A}b_{A}+N_{B}a_{B}b_{B})/L^2$ with $L$ the box dimension]. We use $\delta=0.85$ to model a small size disparity, while the case with $\delta=0.5$ stands for a large size disparity.

We focus on the dynamics of binary hard ellipses with a fixed aspect ratio of $k=2$ ($k=a_A/b_A=a_B/b_B$). The reasons why we choose this aspect ratio are explained as follows. For hard ellipses with large aspect ratio (say $k\geq3$), the system has a strong tendency to form a nematic phase (where particles have their centers of mass at random but exhibit some long-range orientational order) at high densities, although the positional order is highly suppressed~\cite{JCP_139_024501}. We found that this is even true for a binary hard-ellipse mixture with moderate size disparity, at least for small systems composed of several hundred particles. (The formation of a nematic crystal may be avoided for larger systems since the nematic order parameter is expected to decrease with the system size~\cite{Book_deGennes}.) On the other hand, the positional and the orientational order can be simultaneously suppressed in binary hard ellipses with smaller $k$. However, if the aspect ratio is too small (say $k\leq1.5$), the dynamics in the rotational degrees of freedom resembles that of a free rotator, which is quite different from the typical relaxation dynamics in a glass-forming liquid and which has been also observed in MD simulations of hard ellipsoids with aspect ratios close to unity~\cite{PRL_98_265702, EPL_84_16003}. Instead, we have found that a binary hard-ellipse system with an aspect ratio of around $k=2$ exhibits characteristic glassy dynamics on increasing density for the chosen size disparities both translationally and rotationally. We thus focus on a single aspect ratio of $k=2$ in the present work since this will facilitate the analysis of glassy phenomena in binary mixtures. Therefore, the results in the present paper only stand for a particular case of binary hard ellipses. Note that the glass formation in monolayers of monodisperse colloidal ellipsoids with $k\simeq2$ has been studied very recently in experiments~\cite{PRL_110_188301, NC_5_3829} and simulations~\cite{NC_5_3829}. We also mention that the equilibrium phase diagram of monodisperse hard ellipses has been investigated by EDMD simulations~\cite{JCP_139_024501}, and more accurately determined by the recent replica exchange MC simulations~\cite{JCP_140_204502}. For instance, the replica exchange MC simulations~\cite{JCP_140_204502} indicate that an isotropic-plastic transition occurs for monodisperse hard ellipses with aspect ratios up to $k\simeq1.6$, while a nematic liquid crystal forms at sufficiently high densities when $k\geq2.4$. Therefore, the results from the phase diagram of monodisperse hard ellipses support our analysis above that a nematic phase becomes irrelevant for hard ellipses with an aspect ratio of around $k=2$.

We performed EDMD simulations of binary hard ellipses in a square box under periodic boundary conditions. The main ingredients for implementing an EDMD simulation for hard ellipses have been described in Refs.~\cite{JCP_139_024501, CDMD1, CDMD2}. The results in the present paper are presented for the binary mixtures with the total particle number $N=N_A+N_B=500$. Since our aim here is to qualitatively identify the influence of size disparity and composition on the dynamics of binary mixtures, rather than to quantify a true liquid, it is faithful to consider a small system with several hundred particles, where finite size effects for the dynamics are expected to be small~\cite{JCP_139_024501, PRE_86_031502}. We also performed simulations for binary mixtures with $N=300$, and the same conclusion can be drawn from the smaller system. All the particles have the same mass $m$ and the same moment of inertia $I$. Both $m$ and $I$ are set to be unity, with the recognition that the general trends of static and dynamic quantities will not be affected by the choice of $m$ and $I$. The temperature $T$ is irrelevant for athermal systems and remains constant due to the conservation law of the total kinetic energy. We set the temperature as $k_BT=1$. Length and time are expressed in units of $2b_A$ and $\sqrt{4mb_A^{2}/k_{B}T}$. The starting configuration of a binary hard-ellipse system was generated by the Lubachevsky-Stillinger compression algorithm~\cite{LS1, LS2, CDMD1, CDMD2}. At each state point, the system was first equilibrated for at least several relaxation times (see Subsection III B for the definition of relaxation time) before collecting data. We performed at least four independent runs in order to obtain reliable results and improve the statistics.

\section{Results and discussion}

This section begins with a demonstration of the presence of characteristic glassy dynamics on increasing density for the binary hard ellipses in both the translational and the rotational degrees of freedom. We then discuss in detail how size disparity and composition affect the relaxation dynamics of the binary mixtures by investigating the time correlation functions.

\subsection{Characteristic glassy dynamics in the binary hard ellipses}

\begin{figure}[htb]
	\centering
	\includegraphics[angle=0,width=0.45\textwidth]{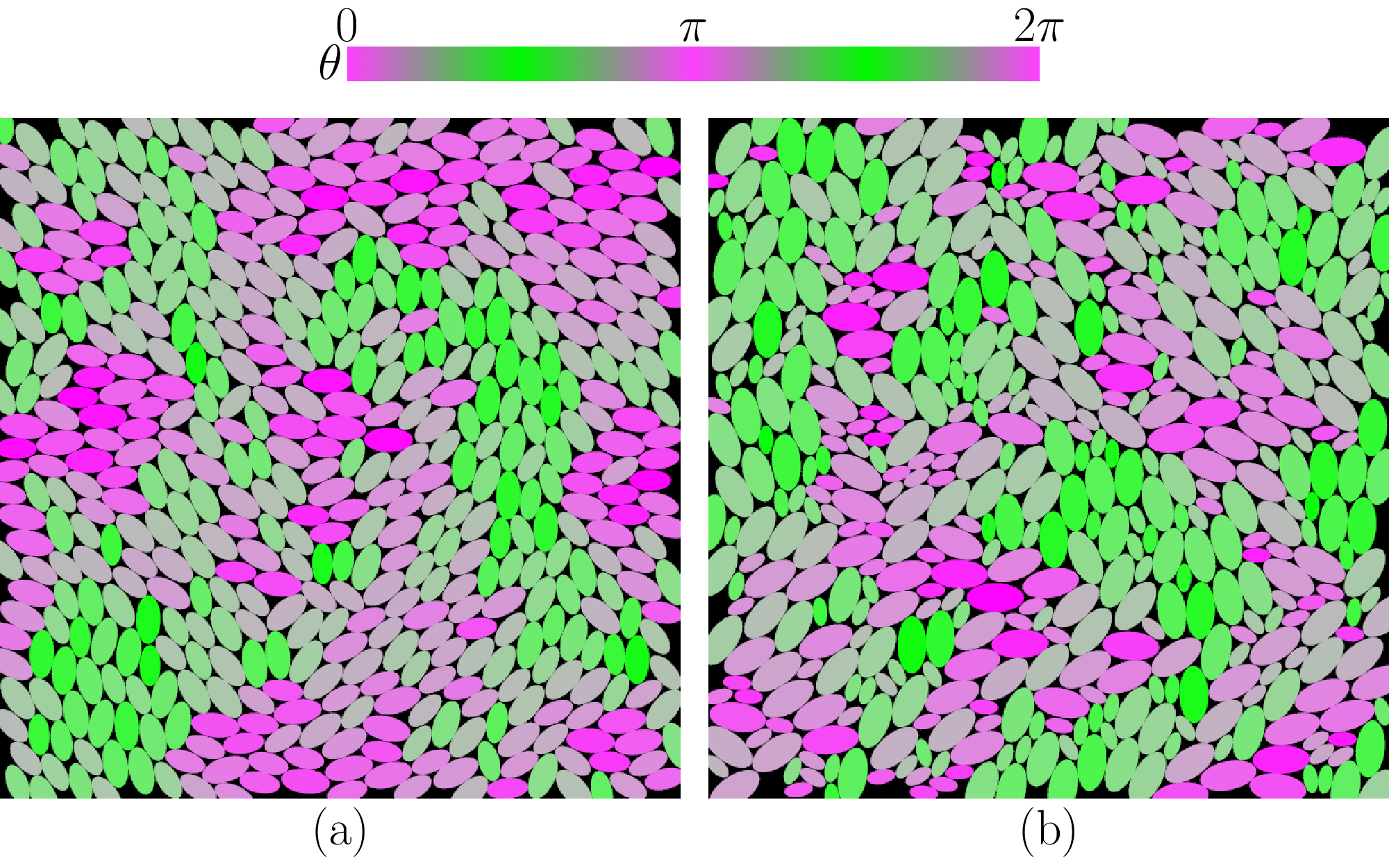}
	\caption{Representative snapshots of the binary hard ellipses with the number concentration $x=0.5$ at a high area fraction of $\phi=0.855$ for (a) $\delta=0.85$ and (b) $\delta=0.5$, respectively. Particles are colored according to their orientation and $\theta$ denotes the angle between the particle semi-major axis and the $x$-axis.}
\end{figure}

We first show that the binary hard-ellipse mixtures with the chosen size disparities exhibit neither long-range positional nor orientational order even at high densities. As an illustration, two representative snapshots after long simulation runs are presented in Fig. 1 for the binary mixtures with $x=0.5$ at a high density $\phi=0.855$ for $\delta=0.85$ and $\delta=0.5$, respectively. Although some short-range order does form at very high densities, the system lacks any long-range order both positionally and orientationally. This can be also confirmed by the pair correlation functions and the angular correlation functions (Figs. S1 and S2 in the ESI). For example, the pair correlation function quickly decays to unity and only displays several peaks at short distances, indicative of the absence of long-range positional order [Figs. S1(a) and S2(a) in the ESI]. Likewise, the angular correlation function decays much faster than that in a nematic liquid crystal even at the highest density studied [Figs. S1(b) and S2(b) in the ESI], and hence, the system remains disordered in the rotational degrees of freedom. Therefore, glass formation occurs upon increasing density for the binary mixtures in both the translational and the rotational degrees of freedom.

\begin{figure}[t]
	\centering
	\includegraphics[angle=0,width=0.45\textwidth]{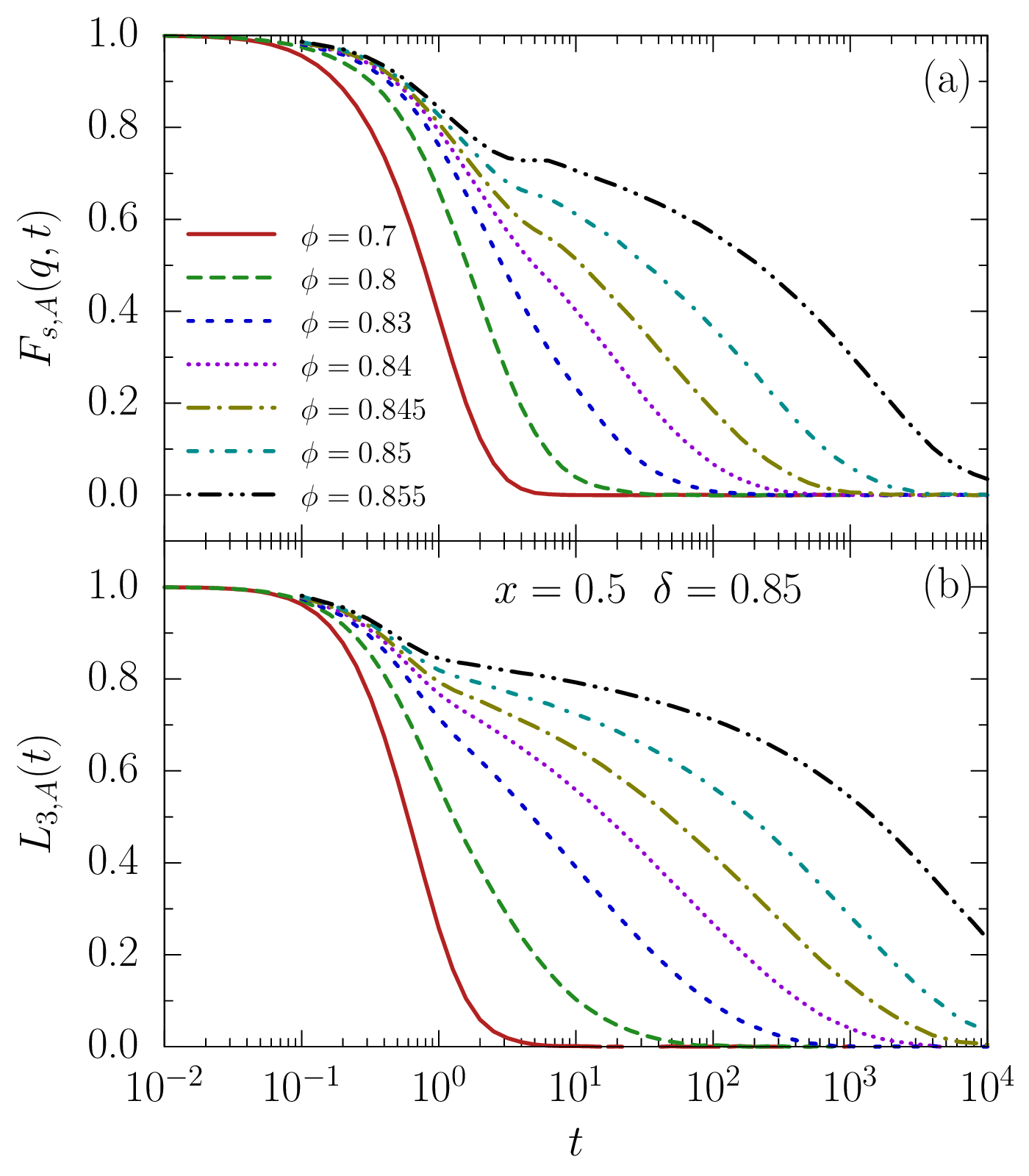}
	\caption{(a) Self-intermediate scattering function $F_{s,A}(q,t)$ at $q\simeq4.0$ and (b) $3$rd order orientational correlation function $L_{3,A}(t)$ for the large particles at various area fractions $\phi$ [indicated in (a)] in the binary hard ellipses with the size ratio $\delta=0.85$ and the number concentration $x=0.5$.}
\end{figure}

\begin{figure}[t]
	\centering
	\includegraphics[angle=0,width=0.45\textwidth]{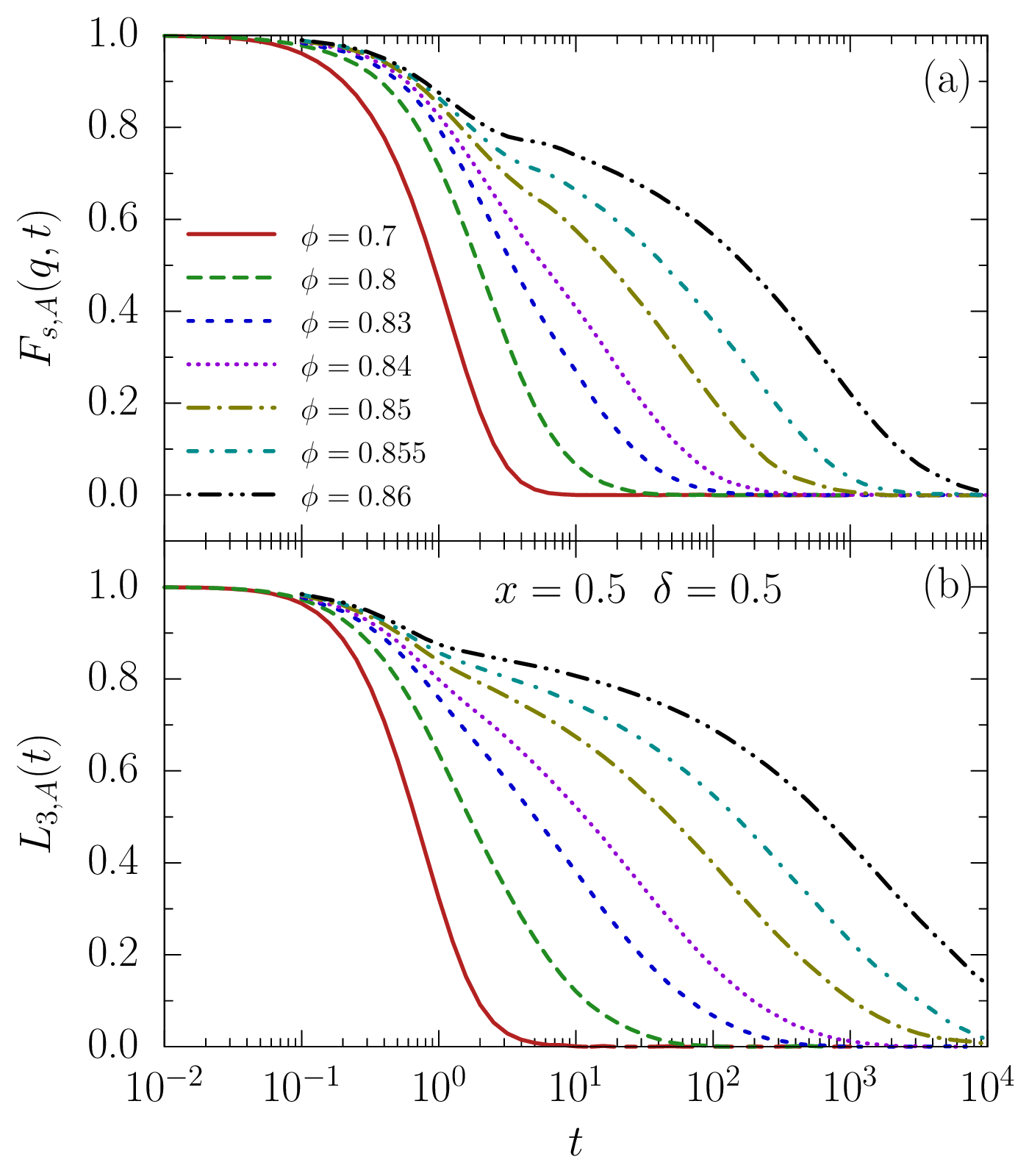}
	\caption{(a) Self-intermediate scattering function $F_{s,A}(q,t)$ at $q\simeq4.0$ and (b) $3$rd order orientational correlation function $L_{3,A}(t)$ for the large particles at various area fractions $\phi$ [indicated in (a)] in the binary hard ellipses with the size ratio $\delta=0.5$ and the number concentration $x=0.5$.}
\end{figure}

The relaxation dynamics we concentrate on here is explored mainly by the self-intermediate scattering function and the $n$th order orientational correlation function, which are defined as
\begin{equation}
	F_{s,\alpha}(q,t)=\frac{1}{N_{\alpha}}<\sum_{j=1}^{N_{\alpha}}\exp\{i\textbf{q}\cdot[\textbf{r}_{\alpha,j}(t)-\textbf{r}_{\alpha,j}(0)]\}>,
\end{equation}
and
\begin{equation}
	L_{n,\alpha}(t)=\frac{1}{N_{\alpha}}<\sum_{j=1}^{N_{\alpha}}\exp\{in[{\theta}_{\alpha,j}(t)-{\theta}_{\alpha,j}(0)]\}>,
\end{equation}
where $\textbf{r}_{\alpha,j}$ and ${\theta}_{\alpha,j}$ are the position and orientation of particle $j$ belonging to species $\alpha$, $<\cdot\cdot\cdot>$ indicates the thermal average, $i=\sqrt{-1}$, $q$ is the wave number, and $n$ is a positive integer. In the following, the results will be given for $F_{s,\alpha}(q,t)$ with $q\simeq4.0$ (which is close to the first peak of the static structure factor for the large particles at high densities), and $L_{n,\alpha}(t)$ with $n=3$, since the reasonable choice of $q$ and $n$ does not affect the qualitative behavior of the corresponding time correlation functions~\cite{PRL_107_065702, PRL_110_188301}. We note, however, that the orientational correlation function has a form different from eqn (2) in three-dimensional anisotropic particle systems and depends crucially on its order $n$, as pointed out in Refs.~\cite{PRE_62_5173, EPL_84_16003, PRE_56_5450}.

$F_{s,\alpha}(q,t)$ is commonly used to characterize the structural relaxation of a supercooled liquid in the translational degrees of freedom, while $L_{n,\alpha}(t)$ has been shown recently to effectively capture the rotational relaxation of a two-dimensional anisotropic system on approaching the glass transition~\cite{PRE_86_041303, PRL_107_065702, NC_5_3829, PRL_110_188301}. Therefore, the glass formation in the binary mixtures upon increasing density can be nicely illustrated by the time correlation functions. Since the variation of the dynamics of both species with density is similar, we only show the results of the time correlation functions for the large particles in Figs. 2 and 3, where we present the results of $F_{s,A}(q,t)$ and $L_{3,A}(t)$ for the binary mixtures with $x=0.5$ at various area fractions for $\delta=0.85$ and $\delta=0.5$, respectively. Both $F_{s,A}(q,t)$ and $L_{3,A}(t)$ are nearly exponential at area fractions below $\phi=0.8$, while they become stretched and develop a two-step relaxation upon further increasing density. The two-step decay is typical of supercooled liquids and reflects the rattling motion of particles being trapped in cages formed by their nearest neighbors at short times ($\beta$-relaxation) and the motion of particles escaping from the cages at long times ($\alpha$-relaxation). In addition, we find that the time correlation functions satisfy the $t$-$\phi$ superposition property, i.e., a master curve can be obtained at long times by plotting $F_{s,\alpha}(q,t)$ or $L_{n,\alpha}(t)$ versus $t/\tau$ at various $\phi$ (Figs. S3 and S4 in the ESI), where $\tau$ is the corresponding relaxation time (see subsection 3.2 for its definition), and that the relaxation times have a power-law dependence on $\phi$ at high densities (see Subsection III B). The above observations are also signatures of the glassy dynamics, as predicted by MCT~\cite{Book_Gotze}. Thus, our results indicate the formation of glasses at high densities in both the translational and the rotational degrees of freedom. Note that the onset density of the two-step decay in the binary hard ellipses is much larger than that in their spherical counterpart (i.e., hard disks); e.g., the two-step decay is already evident at $\phi=0.77$ for a hard-disk system with a polydispersity of $12\%$~\cite{JCP_137_104509}. We will further discuss this point in Subsection III B.

\subsection{Influence of size disparity}

\begin{figure}[t]
	\centering
	\includegraphics[angle=0,width=0.45\textwidth]{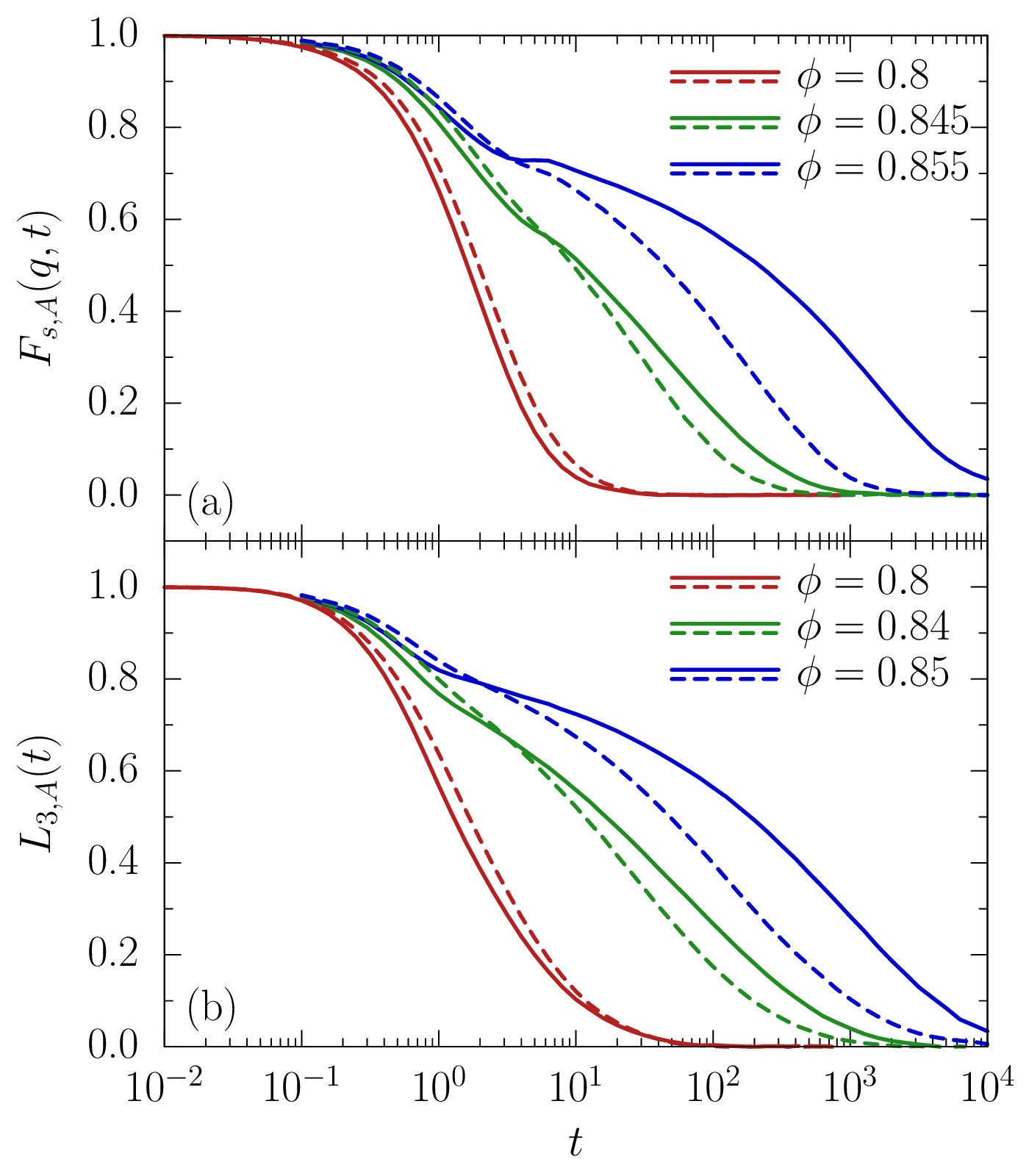}
	\caption{(a) Self-intermediate scattering function $F_{s,A}(q,t)$ at $q\simeq4.0$ and (b) $3$rd order orientational correlation function $L_{3,A}(t)$ for the large particles at three area fractions $\phi$ (indicated in each panel) in the binary hard ellipses with the number concentration $x=0.5$ for $\delta=0.85$ (solid lines) and $\delta=0.5$ (dashed lines).}
\end{figure}

\begin{figure}[t]
	\centering
	\includegraphics[angle=0,width=0.45\textwidth]{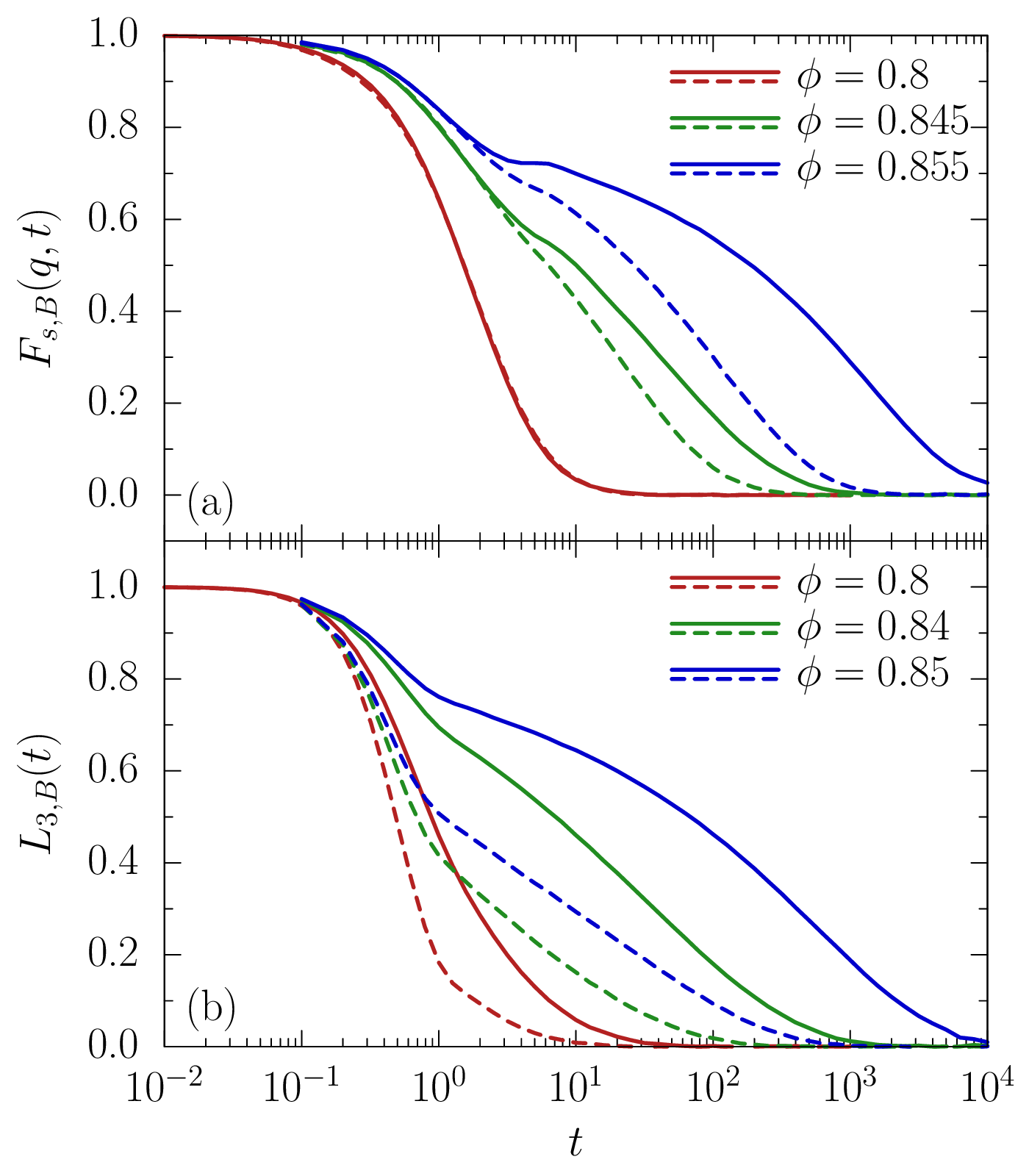}
	\caption{The same as in Fig. 4, but for the small particles.}
\end{figure}

We now discuss the influence of size disparity on the relaxation dynamics. To this end, we directly compare the time correlation functions at different size disparities, when the other parameters are held constant. Figures 4 and 5 display the results of the time correlation functions for both size disparities for fixed $x$ and $\phi$ for $A$ and $B$ particles, respectively. In normal liquid states, the dynamics of binary hard spheres can be quantitatively explained by Enskog's kinetic theory~\cite{PRL_91_085701}, which predicts that an increase of size disparity (smaller $\delta$) leads to a slowing down of the long-time dynamics for the large particles and a speed up for the small particles. Such phenomena are also seen in the binary hard ellipses when the density is low (e.g., compare different types of curves at $\phi=0.8$ in Figs. 4 and 5). Note that the same scenario also holds in the rotational degrees of freedom. When the liquid enters into the supercooled regime, the trend of the dynamics for the small particles remains unchanged (see Fig. 5), since their size always diminishes upon decreasing $\delta$ provided that the large particle size is used as the length unit. While, the influence of size disparity on the relaxation dynamics changes qualitatively for the large particles. In both the translational and the rotational degrees of freedom, we observe that the long-time decay for the large particles becomes faster for smaller $\delta$ (compare different types of curves for $\phi>0.8$ in Fig. 4), a result again similar to that in binary hard spheres~\cite{PRL_91_085701}. This effect becomes more pronounced at higher densities and has been also identified as a remarkable feature of structural relaxation since the behavior is qualitatively different from the Enskog's prediction. 

\begin{figure}[htb]
	\centering
	\includegraphics[angle=0,width=0.45\textwidth]{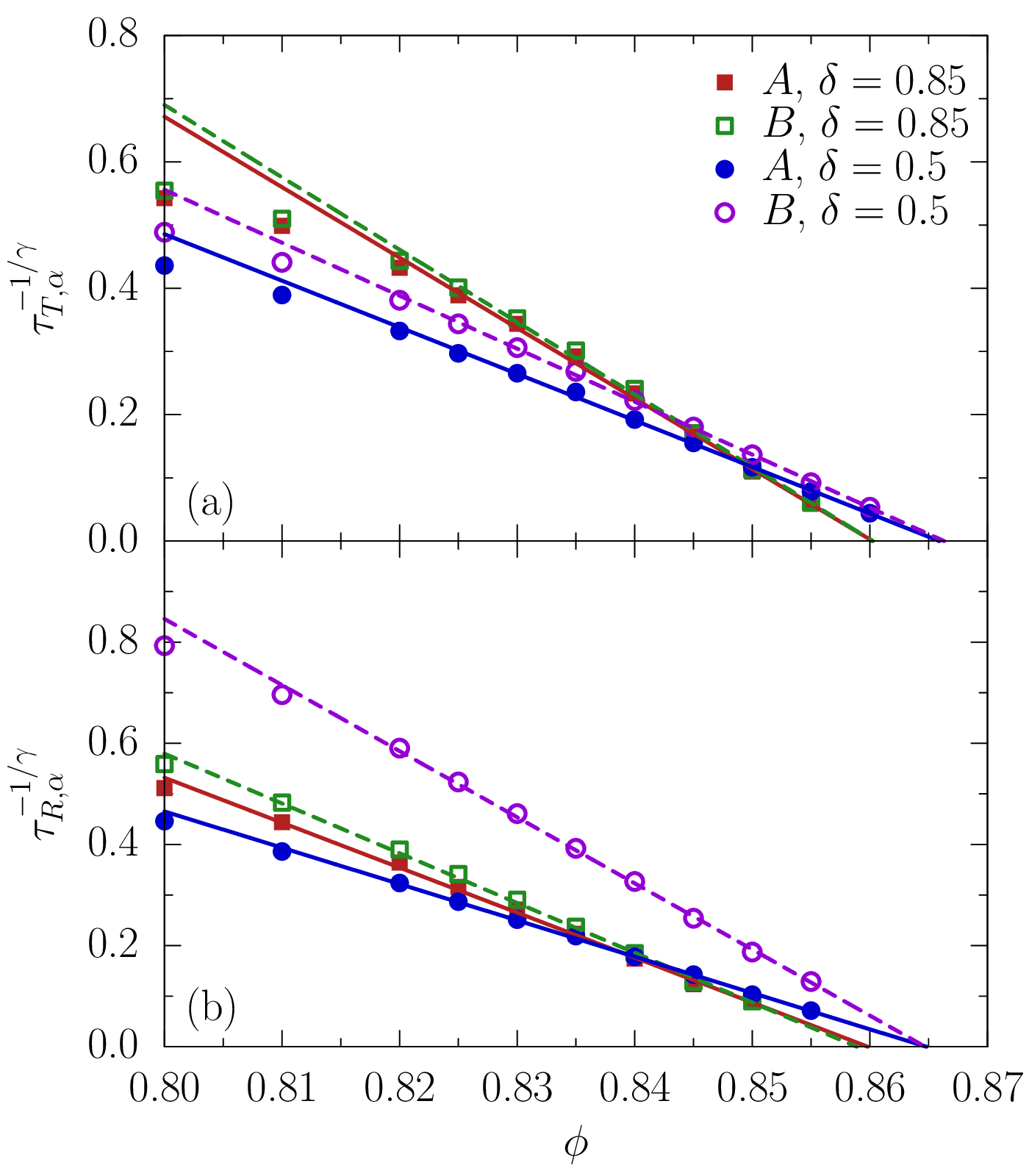}
	\caption{(a) $\tau_{T, \alpha}^{-1/\gamma}$ and (b) $\tau_{R, \alpha}^{-1/\gamma}$ as a function of area fraction $\phi$ for both size ratios. Here, $\tau_{T, \alpha}$ and $\tau_{R, \alpha}$ denote the translational and the rotational relaxation time of particle $\alpha$, respectively. The lines are MCT fits [i.e., $\tau\sim(\phi_c-\phi)^{-\gamma}$] with the fitted results summarized in Table I.}
\end{figure}

\begin{table}[htb]
	\centering
	\small
	\caption{MCT fittings for the binary hard-ellipse mixtures. The table summarizes the results of $\gamma$ and $\phi_{c}$ for both species in both the translational and the rotational degrees of freedom for $\delta=0.85$ and $\delta=0.5$, respectively.}
	\begin{tabular*}{0.48\textwidth}{@{\extracolsep{\fill}}lccccc}
		\hline\hline
		\toprule
		& & \multicolumn{2}{c}{Translational} & \multicolumn{2}{c}{Rotational}\\
		\cmidrule(lr){3-4} \cmidrule(l){5-6}
		$\delta$ & Particle species & $\gamma$ & $\phi_{c}$ & $\gamma$ & $\phi_{c}$\\
		\midrule
		\hline
		$0.85$ & $A$ & $2.94$ & $0.8602$ & $3.47$ & $0.8598$\\
		$0.85$ & $B$ & $2.94$ & $0.8603$ & $3.19$ & $0.8589$\\
		$0.5$   & $A$ & $2.48$ & $0.8659$ & $3.05$ & $0.8648$\\
		$0.5$   & $B$ & $2.47$ & $0.8663$ & $2.70$ & $0.8647$\\
		\bottomrule
		\hline\hline
	\end{tabular*}
\end{table}

The size-disparity effect at high densities, typical of structural relaxation, has further implications. Since the long-time dynamics becomes fast on the approach to the glass transition as the size disparity increases (i.e., as $\delta$ decreases), a higher glass transition density is expected for smaller $\delta$ in both the translational and the rotational degrees of freedom. This conclusion is supported by the empirically determined mode-coupling singularity $\phi_c$, obtained by power-law fits of the $\phi$-dependence of the relaxation time $\tau$, i.e., $\tau\sim(\phi_c-\phi)^{-\gamma}$, where $\phi_c$ and $\gamma$ are fitting parameters. Here, we define the translational relaxation time $\tau_{T, \alpha}$ and the rotational one $\tau_{R, \alpha}$ for particle $\alpha$ as $F_{s,\alpha}(q,t=\tau_{T, \alpha})=0.1$ and $L_{3,\alpha}(t=\tau_{R, \alpha})=0.1$, respectively. As found in other glass formers, both $\gamma$ and $\phi_c$ are sensitive to the range of density fitted, but we confirmed that the general trend of $\phi_c$ in variation with $\delta$ remains independent of the fit interval. Our best fits are shown in Fig. 6 and the fitted results for $\gamma$ and $\phi_c$ are summarized in Table I. Furthermore, the critical area fractions for both types of particles differ by no more than $0.001$ in both the translational and the rotational degrees of freedom, which confirms the coupling of the glassy dynamics for both species. (The fitted exponents $\gamma$ are also similar in our study for different types of particles.) In accord with the influence of size disparity on the relaxation dynamics observed in Figs. 4 and 5, we find that $\phi_c$ for the large particles increases from $0.8602$ to $0.8659$ in the translational degrees of freedom and from $0.8598$ to $0.8648$ in the rotational degrees of freedom, when $\delta$ decreases from $0.85$ to $0.5$. 

The translational glass transition density in the binary hard ellipses is clearly much larger than that in the hard disks, where the MCT critical area fraction is determined to be $\sim 0.791$ when the polydispersity of the system is about $12\%$~\cite{JCP_137_104509}. This observation is in agreement with the numerical studies on the jamming transition in ellipse-shaped particles~\cite{PRE_75_051304, SM_6_2960, PRE_86_041303, NC_5_3829}, which indicate that the jamming transition density in a system of pure ellipses with $k=2$ are significantly higher than that in a system of pure disks. Moreover, we notice that the translational glass transition density $\phi_c^T$ is slightly larger than the rotational one $\phi_c^R$ for both size disparities studied. Nevertheless, they do not differ too much, consistent with the previous analysis based on the self-diffusion of monodisperse hard ellipses~\cite{JCP_139_024501} and recent experiments for monolayers of colloidal ellipsoids with $k\simeq2$~\cite{PRL_110_188301, NC_5_3829}. Of course, the difference between $\phi_c^T$ and $\phi_c^R$ may become fairly evident for large aspect ratios, which is indeed seen in the experiments for monolayers of hard ellipsoids with $k>2$~\cite{PRL_107_065702, NC_5_3829}.

\subsection{Influence of composition}

\begin{figure}[htb]
	\centering
	\includegraphics[angle=0,width=0.45\textwidth]{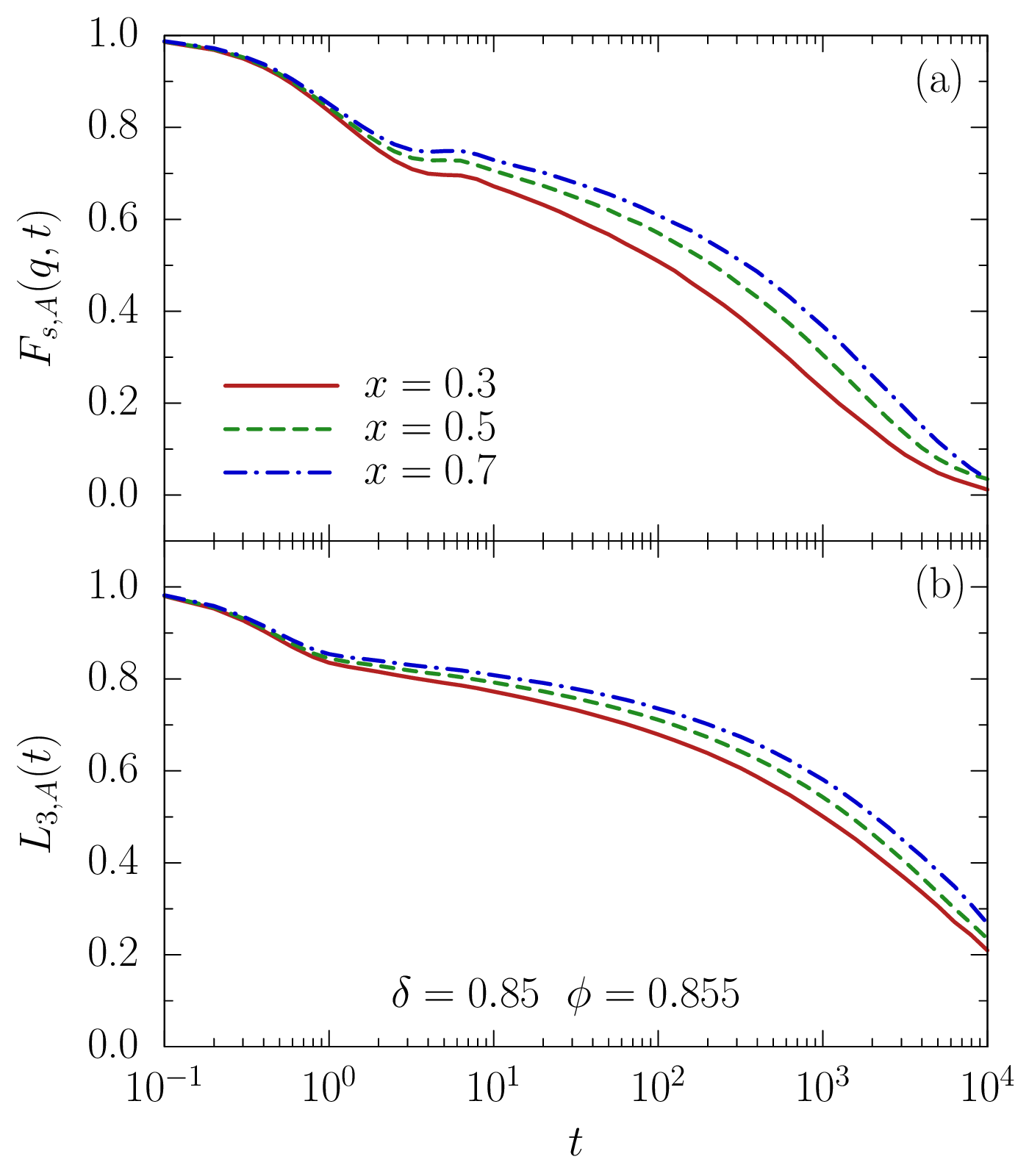}
	\caption{(a) Self-intermediate scattering function $F_{s,A}(q,t)$ at $q\simeq4.0$ and (b) $3$rd order orientational correlation function $L_{3,A}(t)$ for the large particles for various compositions $x$ [indicated in (a)] in the binary hard ellipses with the size ratio $\delta=0.85$ at the area fraction $\phi=0.855$. The results are similar for the small particles.}
\end{figure}

\begin{figure}[htb]
	\centering
	\includegraphics[angle=0,width=0.45\textwidth]{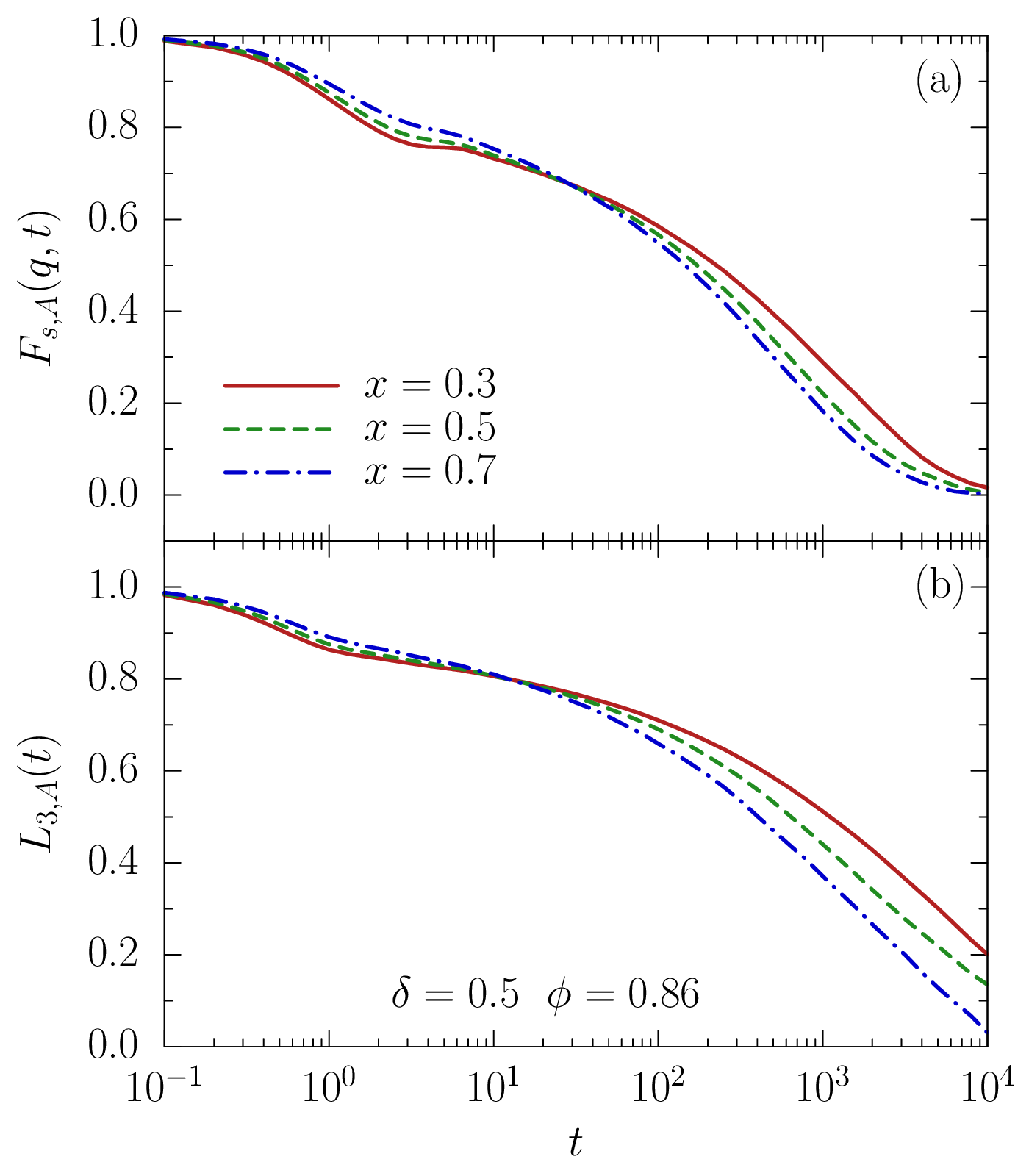}
	\caption{(a) Self-intermediate scattering function $F_{s,A}(q,t)$ at $q\simeq4.0$ and (b) $3$rd order orientational correlation function $L_{3,A}(t)$ for the large particles for various compositions $x$ [indicated in (a)] in the binary hard ellipses with the size ratio $\delta=0.5$ at the area fraction $\phi=0.86$. The results are similar for the small particles.}
\end{figure}

Turning to the influence of mixture's composition on the relaxation dynamics, we focus here on how the number concentration of the small particles affects the time correlation functions at a single high density ($\phi=0.855$ for $\delta=0.85$ and $\phi=0.86$ for $\delta=0.5$), since the previous work~\cite{PRL_91_085701, PRE_80_021503, PRE_83_041503} suggests that  the influence of mixture's composition becomes more pronounced as the liquid is more supercooled. Figures 7 and 8 display the results of $F_{s,A}(q,t)$ and $L_{3,A}(t)$ at a fixed density for various compositions $x$ for $\delta=0.85$ and $\delta=0.5$, respectively. It is obvious that the influence of composition is similar for both $F_{s,A}(q,t)$ and $L_{3,A}(t)$, regardless of whether the size disparity is small or large (compare both panels in Figs. 7 and 8), indicating that mixture's composition affects the translational and the rotational relaxation dynamics in the same way. Moreover, two qualitatively different scenarios, reminiscent of those in binary hard spherical particles, appear in the binary hard-ellipse mixtures, depending on whether the size disparity is small or large. For small size disparity ($\delta=0.85$), a slowing down of the relaxation dynamics occurs for both the initial and the final decay when the number concentration $x$ grows from $0.3$ to $0.7$ at a fixed area fraction, and thus no crossing of the curves in the entire time interval is observed (see Fig. 7). This implies that mixing yields a slight extension of the glass regime in both the translational and the rotational degrees of freedom. By contrast, an increase of $x$ in the binary mixture with large size disparity ($\delta=0.5$) leads to a slowing down of the initial decay but a speed up of the final decay, resulting in the curves' crossing in the time interval for the second relaxation step (see Fig. 8). In this case, mixing induces a plasticization effect and leads to the stabilization of the liquid. As explained in Ref.~\cite{PRE_64_041502}, along with Refs.~\cite{JR_36_947, PRL_68_1422}, this plasticization effect arises because admixing small particles while maintaining volume fraction increases packing efficiency of the system, and thus enhances flow and structural relaxation. Therefore, the influence of composition in binary hard ellipses is qualitatively the same as in binary hard spherical particles. As introduced in Section I, these mixing effects are characteristic of structural relaxation. Our results thus suggest that a universal mechanism for the influence of size disparity and composition on the structural relaxation of binary mixtures can be established in both isotropic and anisotropic particle systems. 

Finally, it would be interesting to explore the dynamic heterogeneity (i.e., the relaxations of particles are not uniform in space) in the binary hard ellipses, since the recent experiments and simulations~\cite{PRL_107_065702, NC_5_3829} reveal that dynamic heterogeneity exhibits remarkable structural features in monolayers of colloidal ellipsoids. For instance, the distributions of translational and rotational fast-particle clusters are found to be anticorrelated in space in monolayers of monodisperse colloidal ellipsoids with $k\simeq6$, and the fasted particles in the translational degrees of freedom form a few ribbonlike clusters aligned with their long axes within the pseudonematic domains, while the clusters of the rotationally fastest particles have branchlike structures extending over several small domains around the domain boundaries. Such topics are absolutely a worthwhile line of research to pursue in the binary hard ellipses. However, the system size used in the present work is not faithful to address such questions. Thus, an analysis of the dynamic heterogeneity in binary mixtures of hard ellipses is not performed in the present paper.

\section{Summary}

In summary, we have shown via EDMD simulations that the glass transition can occur both translationally and rotationally in binary mixtures of hard ellipses with an aspect ratio of $k=2$ and moderate size disparity. For this aspect ratio, the rotational glass transition is found to set in at a density close to the translational one for both size disparities studied. We have examined the influence of size disparity and mixture's composition on the relaxation dynamics in order to test whether or not the binary hard-ellipse mixtures exhibit the dynamical features observed in binary systems composed of spherical particles. Our results indicate that increasing size disparity leads to a speed up of the long-time relaxation dynamics in both the translational and the rotational degrees of freedom. Thus, both translational and rotational glass transition are at higher densities for the binary mixture with larger size disparity. When the number concentration of the small particles is increased, the time evolution of both translational and rotational relaxation dynamics displays two qualitatively different scenarios, depending on whether the size disparity is small or large. For small size disparity, both the initial and the final part of the structural relaxation slow down. While, the short-time dynamics still slows down but the final decay speeds up in the binary mixture with large size disparity. Therefore, the influence of mixture's composition on the relaxation dynamics is also similar to that in binary hard spherical particles. Our results suggest a universal mechanism for the influence of size disparity and mixture's composition on the structural relaxation in both isotropic and anisotropic particle systems. 

\begin{acknowledgments}
This work is supported by the National Basic Research Program of China (973 Program, 2012CB821500), and the National Natural Science Foundation of China (21474111, 21222407) programs.
\end{acknowledgments}

\bibliography{aps}

\pagebreak
\widetext
\clearpage
\begin{center}
\textbf{\large Electronic Supplementary Information}
\end{center}
\setcounter{equation}{0}
\setcounter{figure}{0}
\setcounter{table}{0}
\setcounter{page}{1}
\makeatletter
\renewcommand{\theequation}{S\arabic{equation}}
\renewcommand{\thefigure}{S\arabic{figure}}
 
\textbf {S1 Static structure of the binary hard ellipses}---This section provides additional evidence that the binary hard-ellipse mixtures investigated in the present work exhibit neither long-range positional nor orientational order upon increasing density. Figures S1 and S2 display the pair correlation functions $g_{A}(r)$ and the angular correlation functions $g_{2,A}(r)$ for the large particles at various area fractions $\phi$ for $\delta=0.85$ and $\delta=0.5$, respectively, when the number concentration of the small particles is fixed at $x=0.5$. Definitions for pair correlation function and angular correlation function can be found in Ref.~\cite{JCP_139_024501}. As shown in Figs. S1(a) and S2(a), the pair correlation function $g_{A}(r)$ quickly decays to unity and only displays several peaks at short distances, indicative of the absence of long-range positional order. In addition, Figures S1(b) and S2(b) imply that the angular correlation function $g_{2,A}(r)$ decays much faster than that in a nematic liquid crystal, where the angular correlation function decays slower than the power law $g_{2}(r)\sim r^{-1/4}$~\cite{JCP_139_024501}. Thus, the binary mixtures also remain disordered in the rotational degrees of freedom. Therefore, Figures S1 and S2 further confirm that the long-range order in the binary hard ellipses investigated in the present work is suppressed in both the translational and the rotational degrees of freedom.\\

 \begin{figure}[htb]
 	\centering
 	\includegraphics[angle=0,width=0.9\textwidth]{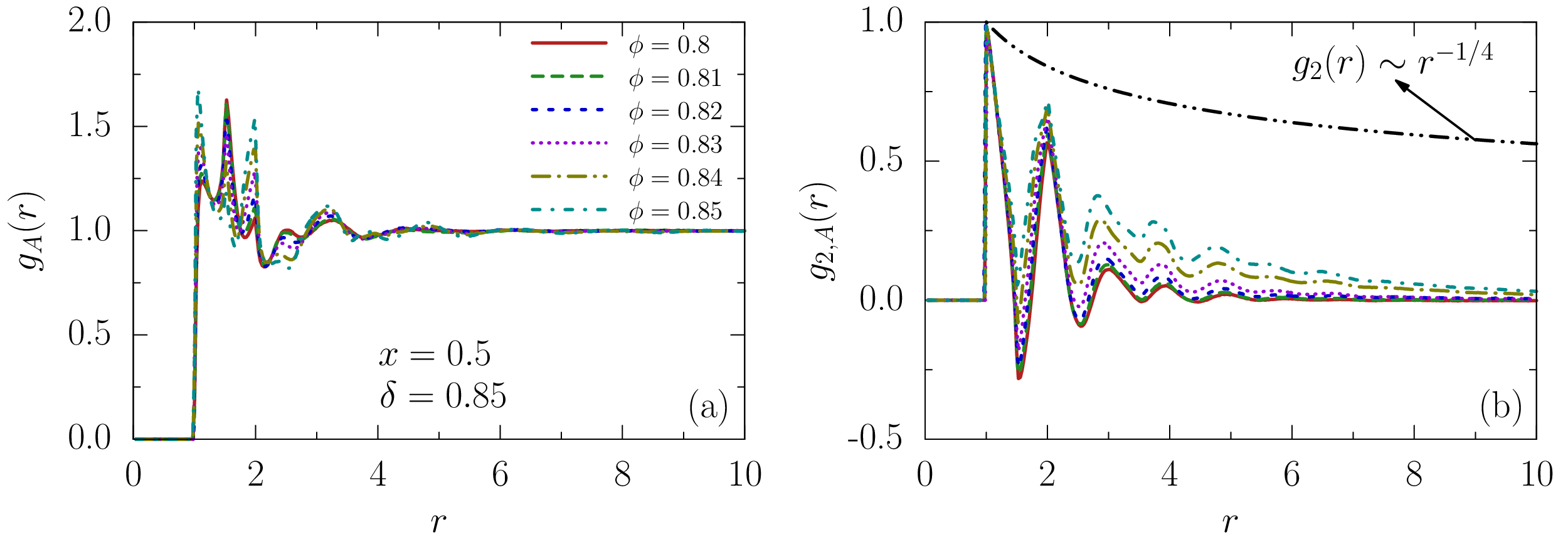}
 	\caption{(a) Pair correlation function $g_{A}(r)$ and (b) angular correlation function $g_{2,A}(r)$ for the large particles at various area fractions $\phi$ [indicated in (a)] in the binary hard ellipses with the size ratio $\delta=0.85$ and the number concentration of the small particles $x=0.5$. The black line in (b) highlights the fact that $g_{2,A}(r)$ for the binary mixture decays much faster than the power law $g_{2}(r)\sim r^{-1/4}$, indicating the absence of any long-range orientational order in the system.}
 \end{figure}
 
 \begin{figure}[htb]
 	\centering
 	\includegraphics[angle=0,width=0.9\textwidth]{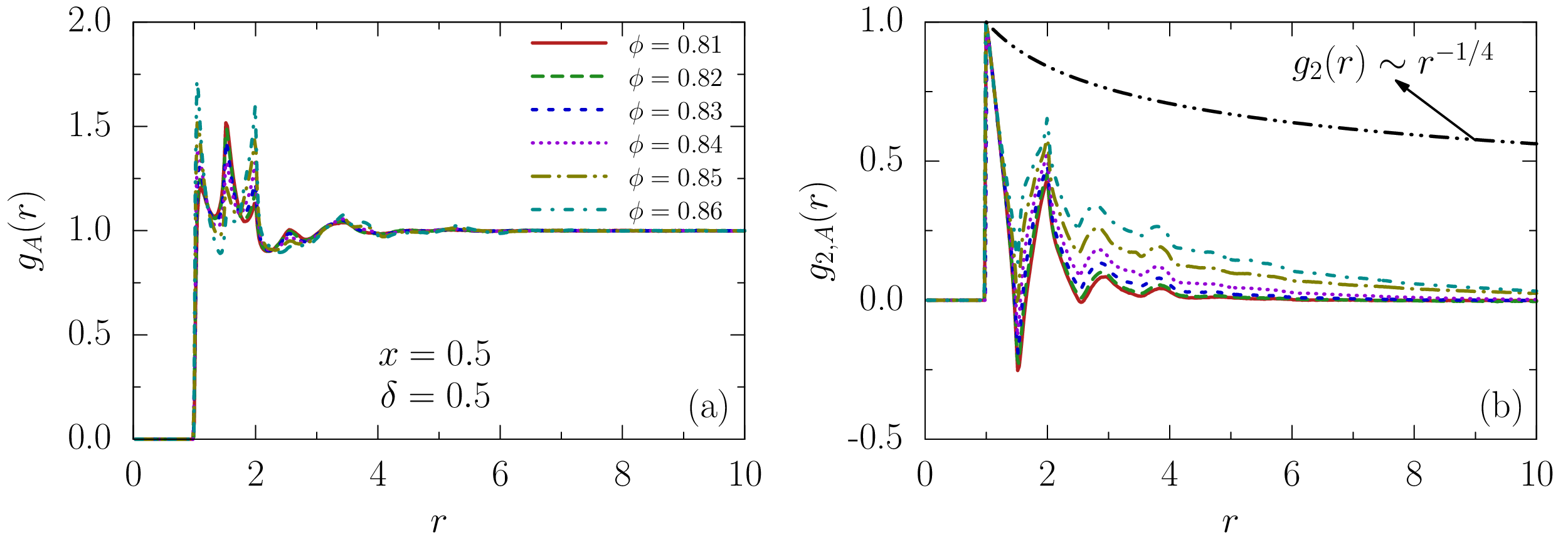}
 	\caption{(a) Pair correlation function $g_{A}(r)$ and (b) angular correlation function $g_{2,A}(r)$ for the large particles at various area fractions $\phi$ [indicated in (a)] in the binary hard ellipses with the size ratio $\delta=0.5$ and the number concentration of the small particles $x=0.5$. The black line in (b) highlights the fact that $g_{2,A}(r)$ for the binary mixture decays much faster than the power law $g_{2}(r)\sim r^{-1/4}$, indicating the absence of any long-range orientational order in the system.}
 \end{figure}

\textbf {S2 Time-density superposition property of time correlation functions in the binary hard ellipses}---This section provides additional results for illustrating the characteristic glassy dynamics in the binary hard ellipses investigated in the present work. Specifically, Figures S3 and S4 display the self-intermediate scattering function $F_{s,A}(q,t/\tau)$ at $q\simeq4.0$ and the $3$rd order orientational correlation function $L_{3,A}(t/\tau)$ for the large particles as a function of $t/\tau$ for $\delta=0.85$ and $\delta=0.5$, respectively. Here, $\tau$ is the relaxation time, defined as the time where the corresponding time correlation function decays to $0.1$. Data collapse is observed at long times for both size ratios and for both $F_{s,A}(q,t/\tau)$ and $L_{3,A}(t/\tau)$, implying that the $t$-$\phi$ superposition property is satisfied in both the translational and the rotational degrees of freedom. This superposition property is also a signature of the glassy dynamics, as predicted by the mode-coupling theory~\cite{Book_Gotze} and observed in other glass-forming liquids (see, e.g., Ref.~\cite{PRE_69_011505}), thereby providing additional evidence for the appearance of characteristic glassy dynamics in the binary hard ellipses in both the translational and the rotational degrees of freedom. We note that the $t$-$\phi$ superposition property in the binary hard ellipses appears only at sufficiently high densities (i.e., $\phi$ is larger than $\sim0.83$ for the binary mixtures investigated), which are typically above the onset density where the time correlation function begins to develop a two-step decay.

\begin{figure}[htb]
	\centering
	\includegraphics[angle=0,width=0.9\textwidth]{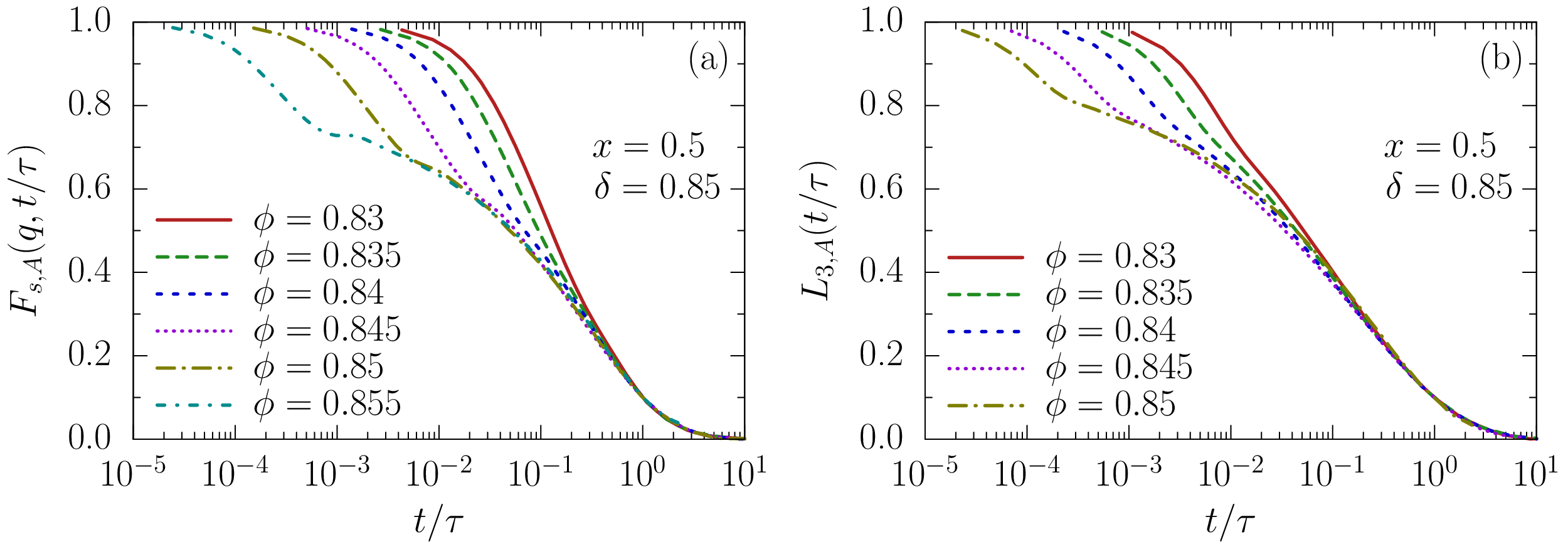}
	\caption{(a) Self-intermediate scattering function $F_{s,A}(q,t/\tau)$ at $q\simeq4.0$ and (b) $3$rd order orientational correlation function $L_{3,A}(t/\tau)$ for the large particles as a function of $t/\tau$ with $\tau$ being the corresponding relaxation time at various area fractions $\phi$ in the binary hard ellipses with the size ratio $\delta=0.85$ and the number concentration of the small particles $x=0.5$. Data collapse at long times indicates that time correlation functions satisfy the $t$-$\phi$ superposition property in the binary hard ellipses in both the translational and the rotational degrees of freedom.}
\end{figure}

\begin{figure}[htb]
	\centering
	\includegraphics[angle=0,width=0.9\textwidth]{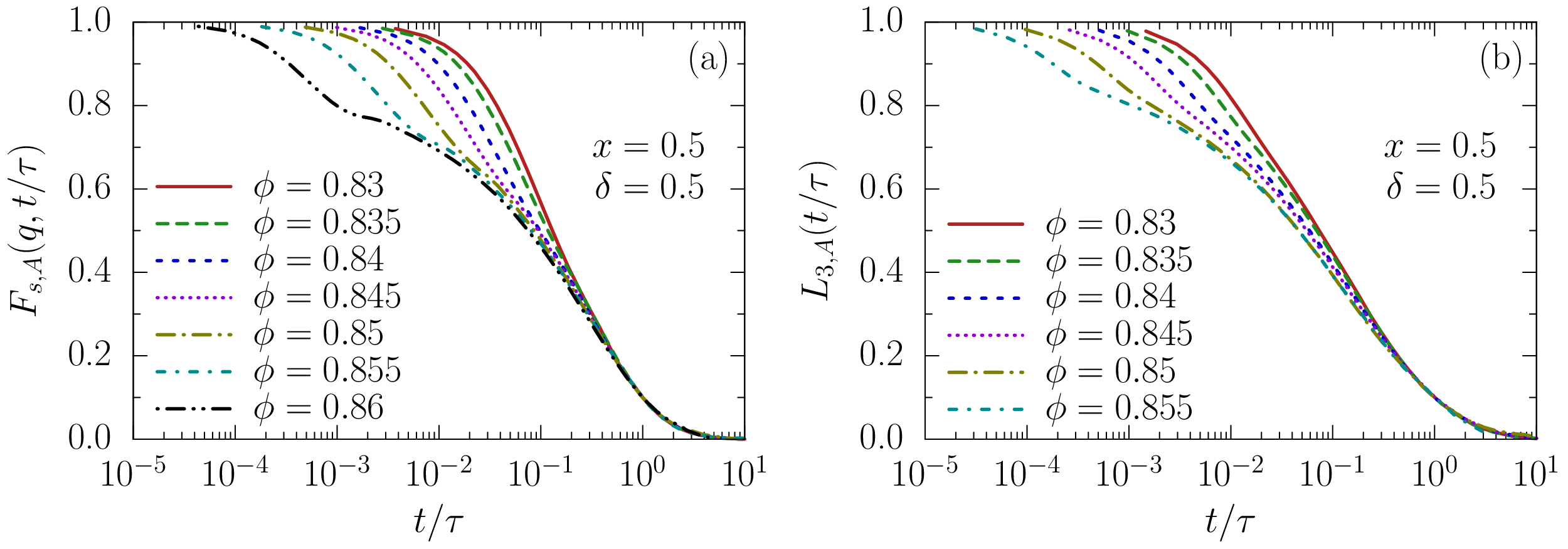}
	\caption{(a) Self-intermediate scattering function $F_{s,A}(q,t/\tau)$ at $q\simeq4.0$ and (b) $3$rd order orientational correlation function $L_{3,A}(t/\tau)$ for the large particles as a function of $t/\tau$ with $\tau$ being the corresponding relaxation time at various area fractions $\phi$ in the binary hard ellipses with the size ratio $\delta=0.5$ and the number concentration of the small particles $x=0.5$. Data collapse at long times indicates that time correlation functions satisfy the $t$-$\phi$ superposition property in the binary hard ellipses in both the translational and the rotational degrees of freedom.}
\end{figure}

\end{document}